# Photoluminescence-driven Broadband Transmitting Directional Optical Nanoantennas

*Kel-Meng See[1], Fan-Cheng Lin[1], Tzu-Yu Chen[1], You-Xin Huang[1], Chen-Hsien Huang[1], A. T. Mina Yesilyurt[2] and Jer-Shing Huang[1,2,3,4*]*

[1] Department of Chemistry, National Tsing Hua University, Hsinchu 30013, Taiwan

[2] Leibniz Institute of Photonic Technology, Albert-Einstein Str. 9, 07745 Jena, Germany

[3] Research Center for Applied Sciences, Academia Sinica, 128 Sec. 2, Academia Road, Nankang District, Taipei 11529, Taiwan

[4] Department of Electrophysics, National Chiao Tung University, Hsinchu 30010, Taiwan

Correspondence should be addressed:
jer-shing.huang@leibniz-ipht.de

**ABSTRACT**

Optical nanoantennas mediate near and far optical fields. Operating a directional nanoantenna in transmitting mode is challenging as the antenna needs to be driven by a nanosized optical-frequency generator, working at the antenna resonance frequency and attached precisely to the antenna feed with a correct orientation. Quantum emitters have been used as optical nanogenerators. However, their precise positioning relative to the nanoantenna is a technical barrier to the practical implementation. One unique source to drive nanoantenna is the photoluminescence of the antenna material as the operation frequency range corresponds to the electronic transitions in matter. Here, we exploit plasmon-modulated photoluminescence (PMPL) as an effective optical source to drive directional nanoantennas. We experimentally realize two technically challenging theoretical proposals, namely, optical nano-spectrometer based on Yagi-Uda nanoantennas and tunable broadband directional emission from log-periodic nanoantennas. Using photoluminescence from the nanoantenna as an optical source promotes practical implementation of transmitting optical nanoantennas.

# Introduction

Optical nanoantennas mediate near and far optical fields and offer opportunities to control light and enhance light-matter interaction [1,2]. Directional nanoantennas are interesting because they promise efficient wireless communication in optical nanocircuits [3-5], provide opportunity to create directional photon sources [6-11] and find various applications in spectroscopic analysis and sensing [12-16]. To gain directionality, the most commonly used design is the Yagi-Uda nanoantenna due to its simplicity, flexibility and robustness [7,8,11,12,17-19]. To achieve tunable broadband directionality, log-periodic nanoantennas [20] and tapered nanoantennas have been proposed [21,22]. It is, however, technically challenging to drive the broadband nanoantennas in transmitting mode because multiple nanosized quantum emitters emitting at different frequencies need to be selectively and precisely attached to specific nanoantenna elements, which are extremely small and separated just by sub-wavelength distances. If dipolar emitters are used, an additional challenge is to align the dipole orientation to effectively excite the longitudinal resonance of the antenna element. A unique optical source to drive a plasmonic nanoantenna is the one-photon photoluminescence from the metal [23-33] of the antenna itself. This unique opportunity finds no radio-frequency (RF) analogs since it is based on the fact that optical nanoantennas operate in the frequency range corresponding to electronic transitions in matter. Typically, bulk metals at room-temperature are not considered as luminescent materials because the quantum efficiency of photoluminescence is very low ($\sim 10^{-10}$) [23,24]. However, as metals are shaped into metallic nanoparticles, localized surface plasmon resonance (LSPR) can greatly enhance the quantum efficiency of photoluminescence to $10^{-4}$ [25,27-32], making photoluminescence bright enough for practical applications [28,33]. Compared to the emission from fluorescent molecules or quantum dots, photoluminescence from metallic nanoparticles is broadband, does not blink or bleach, and is not easily saturated. Moreover, the absorption cross section of metallic nanoparticles and the relaxation rate of photoluminescence are much larger than dye molecules, rendering photoluminescence an attractive optical source to drive transmitting optical nanoantennas.

In this work, we use PMPL of gold as an effective optical source to drive directional nanoantennas and experimentally realize broadband directional emission. PMPL provides stable light source with controllable polarization and tunable emission band, which automatically matches the resonance of corresponding antenna element. We first present morphology-dependent PMPL from antenna elements and show narrow-band unidirectional emission from PL-driven Yagi-Uda nanoantennas. Secondly, we demonstrate color sorting of broadband photoluminescence using a well-designed L-shaped Yagi-Uda (LYU) nanoantenna, realizing Engheta and coworkers' theoretical proposal of Yagi-Uda nanoantenna-based nano-spectrometer [17]. Finally, we experimentally demonstrate tunable directional emission from a broadband log-periodic nanoantenna driven by PMPL. Since PMPL is automatically tuned to the operational frequency of the feed and serves as a perfectly matched optical source, we avoid the major technical challenge and successfully realize the theoretical proposal by van Hulst and coworkers. [20].

## Results

**Plasmon-modulated photoluminescence.** We first examined the plasmon modulation effect on the spectrum, polarization and emission pattern of photoluminescence from gold nanostructures. The photoluminescence was excited by focusing a circularly polarized continuous-wave laser at 532 nm onto gold nanostructures. The emission patterns and spectra of the photoluminescence from a self-assembled gold nanorod and an octahedron are shown in Fig. 1a and 1b, respectively. For the gold nanorod (diameter = 38 nm, length = 72 nm), the photoluminescence is strongly enhanced at 650 nm by the longitudinal LSPR and shows longitudinal polarization as well as clear dipolar emission pattern (Fig. 1a). For a gold octahedron (diameter = 75 nm), the photoluminescence shows relatively small anisotropy in polarization and the emission pattern is rather homogeneous in all directions (Fig. 1b). This confirms that the polarization, spectrum and emission pattern of photoluminescence from gold is strongly modulated by LSPR. Next, we fabricate a series of single-crystalline gold nanorods with nominally identical width (50 nm) and height (40 nm) and various lengths between 80 nm and 140 nm. The fabrication method employs focused-ion beam to mill into chemically synthesized single-crystalline gold flakes [34-37]. Figure 1a shows the modulation effect of LSPR on the photoluminescence spectrum. The peak position in the normalized photoluminescence spectra well follows that of the longitudinal LSPR in the scattering spectra. Increasing aspect ratio of the nanorod not only red shifts the peak position but also decreases the intensity of photoluminescence. This has can be explained by the decreasing coupling strength between the LSPR and electron-hole pairs as the LSPR moves away from the excitation wavelength [30]. For this reason, the LSPR enhanced photoluminescence peak of the 140-nm nanorod at 920 nm is relatively dim and the weak peak around 700 nm stands out in the normalized spectrum. The weak 700-nm peak is attributed to the out-of-plane LSPR because it also appears in the photoluminescence spectrum of unstructured flat area of the same flake and contributes to the omnidirectional background in the back-focal-plane. The experimental results shown in Fig. 1a served as reference for us to design our Yagi-Uda nanoantennas.

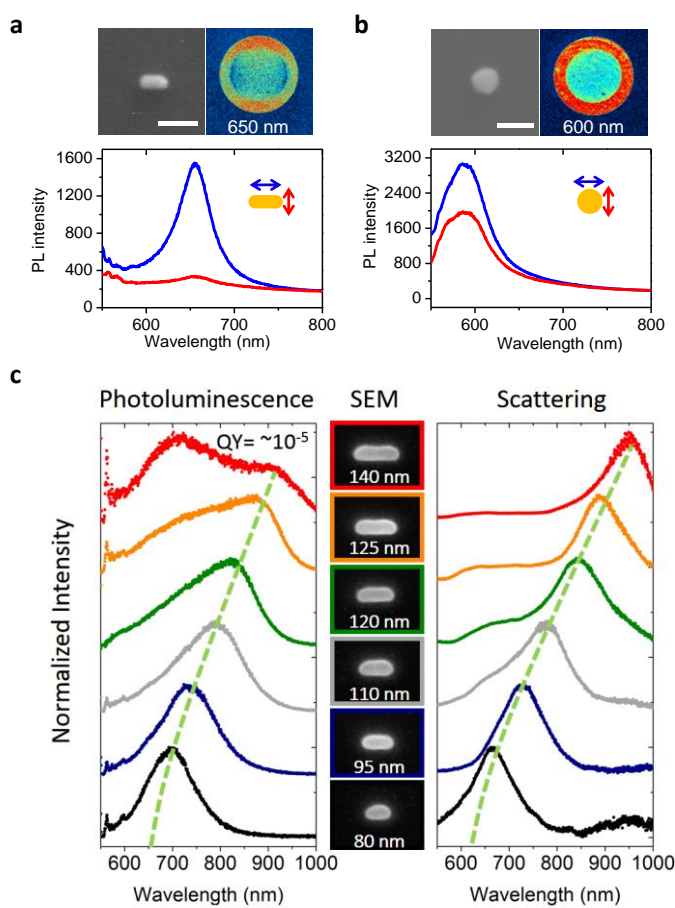

**Figure 1.** a. Top panels show the SEM image of a gold nanorod (length = 72 nm, width = 38 nm) and its back-focal-plane image of the photoluminescence recorded without any polarizer but with a bandpass filters centered at 650 nm. Lower panel shows the photoluminescence spectra recorded with a polarizer orientated in longitudinal (blue) and transverse (red) direction of the nanorod. b. Top panels show the SEM image of a gold octahedron and its back-focal-plane image of the photoluminescence recorded without any polarizer but with a bandpass filters centered at 600 nm. Lower panel is the photoluminescence spectra recorded with a polarizer orientated in two orthogonal directions. Scale bars are 100 nm. c. Normalized photoluminescence spectra (left), SEM images (middle) and dark-field scattering spectra (right) of a series of gold nanorods fabricated by focused-ion beam milling into a single-crystalline gold flake. The dashed lines in the spectra are guides to the eye that track the longitudinal LSPR. The collection polarization is along the long axis of the nanorods.

**Unidirectional emission from PMPL-driven Yagi-Uda nanoantennas.** Having characterized the emission properties of PMPL from single plasmonic elements, we next engineered multiple elements into functional nanoantennas and demonstrated narrow band unidirectional emission from PMPL-driven Yagi-Uda nanoantennas. Yagi-Uda nanoantennas optimized for directional emission at 650 nm and 850 nm were fabricated. Figure 2a shows the SEM images of the Yagi-Uda nanoantennas. SEM images covering large area for all fabricated nanostructures can be found in Fig. S1 in the Supporting Information. Each nanoantenna contains one reflector, one feed and three director elements. The dimensions of the elements and the distance between elements are critical for directionality. In this work, the width and height of each antenna element are kept at 50 nm and 40 nm, respectively, and the free parameters to optimize the directionality are the element length and the inter-element distance. The optimal dimensions are summarized in Fig. S2 (Supporting Information). Real dimensions estimated from SEM images of the fabricated structures are used in the simulations. In addition to the Yagi-Uda nanoantennas, we have also fabricated isolated single reflectors, feeds and directors in order to obtain the emission and scattering spectrum of each element. The simulated scattering cross section and the normalized photoluminescence and dark-field scattering spectra of the single elements of the two Yagi-Uda nanoantennas are shown in Fig. 2b. The dimensions used in the simulations were obtained from the SEM images of the real structures. The photoluminescence spectra were obtained by focusing a circularly polarized 532 nm continuous-wave (CW) laser to the center of the elements and recording the emission spectrum with polarization along the element's long axis. As can be seen, the spectra of the feed elements (black traces in Fig. 2b) show PMPL peaks at the two targeted wavelengths and the resonances of the reflectors and directors are red- and blue-shifted relative to that of the corresponding feed elements, respectively. Dark-field scattering spectra show similar relative spectral shifts. The relative spectral shift between elements is of key importance for Yagi-Uda nanoantennas because it concerns the required phase retardation between elements for coherent interference, which gains the directionality.

To visualize the narrow band unidirectional emission, we illuminated the feed element of the Yagi-Uda nanoantennas with a circularly polarized 532 nm laser spot and record the photoluminescence patterns at the back focal plane of the microscope objective. A series of bandpass filters (bandwidth =

40 nm) centered at 650 nm, 700 nm, 750 nm, 800 nm and 850 nm were used to study the wavelength-dependent directionality. The experimental and simulated back-focal-plane emission patterns are shown in Fig. 2c. At the designed working wavelengths, the back focal plane images show pronounced unidirectional patterns. To quantitatively evaluate the directionality, we calculated the front-to-back (F/B) ratio of each antenna at different wavelengths. The F/B ratio is defined as the logarithmic ratio of the maximum forward intensity ($I_F$) to the backward intensity ($I_B$) [8], *i.e.* F/B ratio $= 10 \log \frac{I_F}{I_B}$.

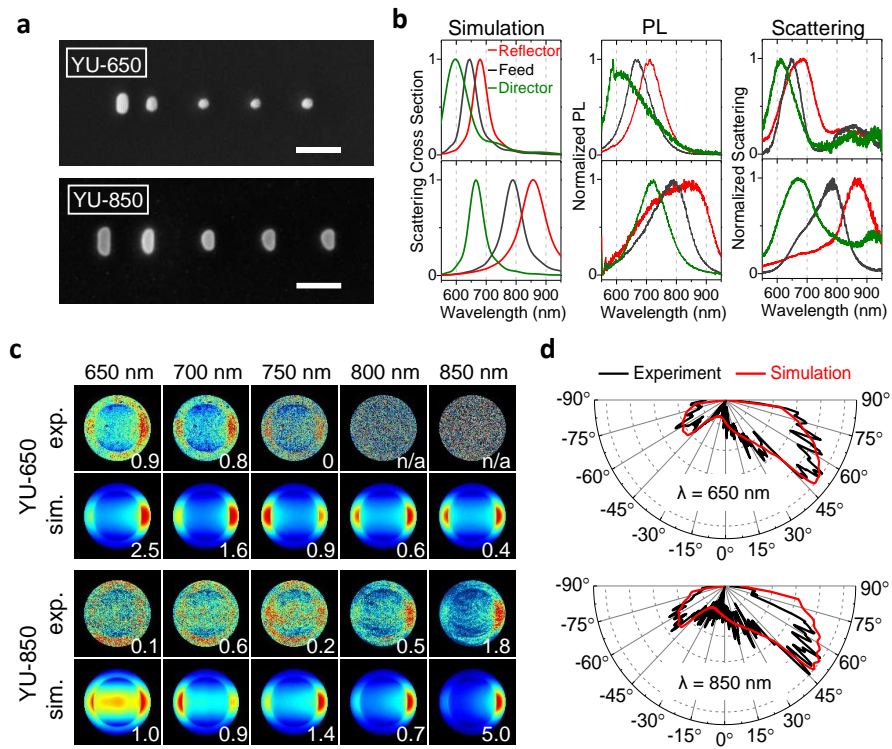

**Figure 2.** a. SEM images of gold Yagi-Uda nanoantennas for directional emission at 650 nm (top panel) and 850 nm (bottom panel). All scale bars are 200 nm. b. Simulated scattering cross section (left column), normalized experimental photoluminescence spectra (middle column) and normalized experimental scattering spectra of the reflector (red traces), feed (black traces) and director (green traces) of the corresponding Yagi-Uda nanoantennas. c. Experimental (upper rows) and simulated (lower rows) back-focal-plane emission patterns recorded within a narrow band centered at 650 nm, 700 nm, 750 nm, 800 nm and 850 nm, corresponding to the columns from the left to the right side. The F/B ratios are marked on each image. The intensity scale has been optimized for each image to best reflect the directionality. d. Normalized experimental (black traces) and simulated angular radiation patterns (red traces) of the photoluminescence from the corresponding Yagi-Uda nanoantennas.

The Yagi-Uda nanoantenna designed for 650 nm (YU-650) shows the highest F/B ratio of 0.9 at 650 nm and the photoluminescence intensity drops to undetectable level in the spectral regime beyond 800 nm. For Yagi-Uda nanoantennas optimized for 850 nm (YU-850), the best F/B ratio is 1.8 at 850 nm. Compared to the undetectable emission pattern of YU-650 at 850 nm, YU-850 apparently is a brighter photon source for 850 nm. This is a direct proof that that the LSPR of the nanostructure can greatly boost the efficiency of photoluminescence [25, 27-32]. Figure 2d shows the angular radiation patterns of the photoluminescence from the two Yagi-Uda nanoantennas at their working wavelengths. Both antennas show clear unidirectional angular pattern and the experimental results agree well with the simulated ones. The photoluminescence is guided into the half space of the substrate because of the relatively high index of ITO glass. In-plane directional emission can be achieved by immersing the antennas into homogeneous medium [19] or using index matching medium [4].

**Yagi-Uda nanoantenna-based nano-spectrometer.** In the following, we show the first experimental realization of a nano-spectrometer based on dual-band Yagi-Uda nanoantennas. This idea was originally proposed by Li *et al.* and Engheta [17] but has not been experimentally realized yet. The main difficulty is to have a suitable broadband optical nanogenerator precisely positioned to simultaneously drive two Yagi-Uda nanoantennas pointing to different directions. Since photoluminescence from gold is broadband and completely depolarized, photoluminescence from a common gold feed element is a perfect photon source to realize this theoretical proposal. To this end, we designed and fabricated an "L-shaped" Yagi-Uda (LYU) nanoantenna, which contains two Yagi-Uda nanoantennas optimized for two wavelengths, pointing to two orthogonal in-plane directions and sharing a common feed element, as shown in Fig. 3a. The in-plane azimuthal directions are indicated with four primary cardinal directions, namely north (N), east (E), west (W) and south (S). The Yagi-Uda antenna pointing to the north is optimized for directional emission at 850 nm, while the other one pointing to the east is designed for 700 nm. The common feed element is an ellipsoid with longitudinal and transverse resonance matching the working wavelengths of the two guiding branches. Figure 3b shows the wavelength-dependent emission patterns recorded at the back focal plane of the microscope objective. A polarizer was used to select the polarization and thus the emission from one of the two branches, *i.e.* the E-W polarization for the north branch and N-S polarization for the east one. For E-W polarization, the photoluminescence shows the highest F/B ratio of 1.4 at 850 nm pointing to the north. For N-S polarization, the photoluminescence show maximum directionality (F/B

= 1.1) to the east at 700 nm. To better visualize the effect of wavelength-dependent directionality, we plotted the F/B ratios in the primary cardinal directions as a function of the wavelength in Fig. 3c. The LYU successfully sorts the emission at 850 nm to the north and 700 nm to the east. This experiment realizes the theoretical proposal of nano-spectrometer based on narrowband Yagi-Uda nanoantennas [17].

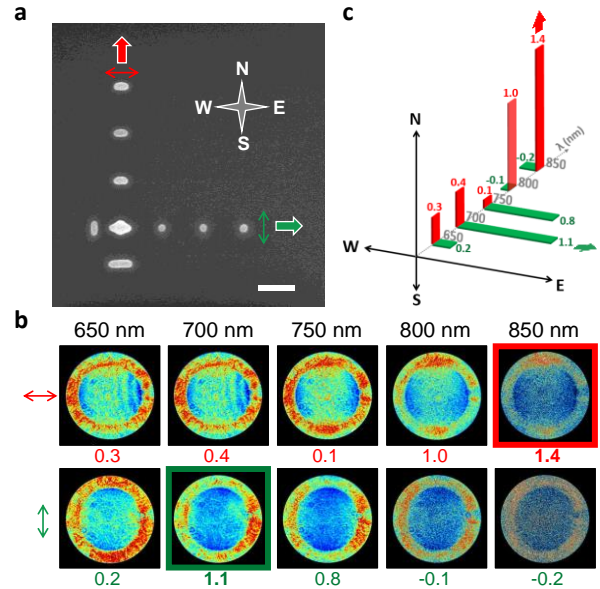

**Figure 3.** a. SEM image of a dual-color L-shaped Yagi-Uda nanoantenna designed for directing emission at 850 nm to the north (red arrow) and 700 nm to the east (green arrow). In-plane directions are indicated by primary cardinal directions, namely north (N), east (E), west (W) and south (S). The double arrows indicate the polarization direction of the collected PL. Scale bar is 200 nm. b. Experimental back-focal-plane images of the photoluminescence recorded with E-W polarization (red double arrow, upper row) and N-S polarization (green double arrow, lower row). F/B ratios are indicated below each image. The intensity scale for all photoluminescence patterns is the same. c. In-plane photoluminescence directionality expressed as a histogram of F/B ratios with respect to the wavelength. Collection polarization at E-W and N-S direction is indicated by red and green color, respectively. The F/B ratio is defined to be positive in the north and the east direction.

**Broadband unidirectional emission from log-periodic nanoantennas.** Finally, we demonstrate the first experimental realization of tunable broadband unidirectional emission from log-periodic nanoantennas (LPNAs). Using LPNAs, directional emission at any desired wavelengths within a broad wavelength window can be achieved by selectively driving antenna elements at their corresponding frequencies. Until now, transmitting LPNAs have not been realized mainly due to the technical barrier that multiple optical nanogenerators need to be precisely attached to specific feed elements in order to drive the elements at the correct optical frequencies. This barrier can be readily overcome by using PMPL as the driving source because the polarization and spectrum of photoluminescence are tuned by the LSPR to perfectly match that of the antenna element.

Experimentally, one can easily tune the wavelength of the directional emission by selectively exciting photoluminescence of different elements. To design effective LPNAs, important design parameters include the length ($a_n$) and width ($b_n$) of the $n^{th}$ element as well as the distance to the next element ($d_n$). The dimensions of elements and the inter-element distance shrink gradually with a fixed scaling factor $r$ [20]. Once $a_n$, $b_n$, $d_n$ and $r$ are chosen, the geometrical parameters of all elements can be determined and the geometrical parameters of two adjacent elements should follow $r = a_{n+1}/a_n = b_{n+1}/b_n = d_{n+1}/d_n$. The available bandwidth is determined by the spectral window defined by the resonance of the largest and the smallest elements. In this work, we have fabricated the LPNAs with $r = 1.2$, $a_1 = 69$ nm, $b_1 = 43$ nm, and $d_1 = 117$ nm to address the spectral range between 650 nm and 850 nm.

Figure 4a shows the SEM image of one exemplary LPNA and the corresponding photoluminescence map recorded by raster scanning the sample position relative to the excitation focal spot of the 532 CW laser. The photoluminescence was collected by the same objective followed by a dichroic mirror to reject the excitation light and aligned onto an avalanche photodiode for photoluminescence maps, the entrance slit of a spectrometer for photoluminescence spectra, or a CCD at the back focal plane of the objective to obtain the emission pattern. With the photoluminescence map, we can choose to precisely excite individual elements, as marked in Fig. 4a. The photoluminescence spectra of solitary antenna elements are presented in Fig. 4b. The peak positions in photoluminescence spectra red shifts from about 650 nm to 750 nm with increasing element dimensions. By focusing the laser focal spot on element # 2 to #5, we were able to selectively drive the LPNA at different wavelengths between 650 nm to 750 nm and tune the directional emission to the desired wavelength. The experimental back-focal plane images of the photoluminescence recorded at various wavelengths are summarized in Fig. 4c with their corresponding F/B ratios. As the excitation focal spot moves from element #2 to #5, the maximum directionality shifts to longer wavelengths. The wavelength for the highest F/B ratio coincides with the peak position of the photoluminescence spectra of single elements shown in Fig. 4a. Overall, the LPNA offers directional emission at the wavelength that can be tuned within the spectral range between 650 nm and 850 nm.

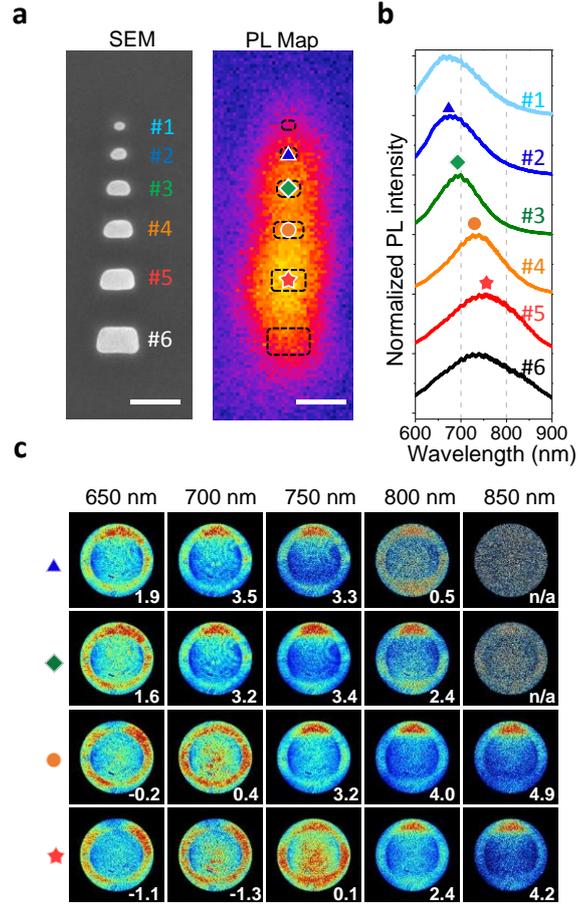

**Figure 4.** a. SEM image (left) and photoluminescence map (right) of a log-periodic nanoantenna. The elements are marked with #1 to #6 for the convenience of indication. The photoluminescence map was recorded by raster scanning the sample and recording the photoluminescence intensity. Excitation focused on element #2 to #5 indicated with blue triangle, green diamond, orange dot and red star, correspondingly. Scale bars are 200 nm. b. Photoluminescence spectra of individual elements. The photoluminescence spectra were obtained by exciting duplicated solitary single elements identical to that in the LPNA. Peak positions of the photoluminescence are marked by the corresponding symbol for excitation positions. c. Experimental back-focal-plane images recorded at various wavelengths under excitation focused on element #2 to #5, as indicated by the symbols. Corresponding F/B ratios are given on each image.

**Discussion**

We present unidirectional emission from PMPL-driven Yagi-Uda nanoantennas. Using photoluminescence from the feed element is a practical way to precisely drive the Yagi-Uda nanoantennas at its working frequency. Based on narrowband Yagi-Uda nanoantennas, we demonstrate a nano-spectrometer based on an L-shaped Yagi-Uda nanoantenna consisting of two orthogonally arranged guiding branches sharing a common feed element. Because the photoluminescence from gold is broadband and completely depolarized, the two sets of Yagi-Uda antennas select and direct the photoluminescence that match their guiding wavelengths. This can be potentially used as a nanoscale analyzer to analyze the emission spectrum of a single emitter sitting at the common feed element. The dual-band Yagi-Uda nanoantenna can also simultaneously operate as receiving antenna in one direction and transmitting antenna in the other direction. In this way, a quantum emitter sitting on the common feed can be effectively excited by the optical power collected wirelessly by the receiving branch and emit the radiation at a different frequency into another direction via the transmitting branch, realizing effective wireless communication between quantum emitters in optical regime [17]. Finally, we exploit the PMPL to achieve the first broadband tunable unidirectional emission from log-periodic nanoantenna. This provides us a simple broadband directional light source with easily tunable wavelength. Using PMPL to drive Yagi-Uda nanoantennas greatly enhances the feasibility of Yagi-Uda nanoantennas as a directional photon source. The possibility of using photoluminescence as an optical driving source marks the fundamental difference between optical nanoantennas and their RF counter parts. This concept is also promising for optical antennas made of other materials, for example, emitting semiconductors [38,39], high-index halide perovskite dielectrics [40] and low-dimensional nanomaterials [41,42] are suitable antenna materials that have much higher quantum efficiency compared to noble metal nanoparticles.

**Methods**

**1. Sample fabrication**

Seed-mediated growth methods were used to chemically synthesize gold nanorods [43] and gold nanooctahedra [44]. Nanoantennas and antenna elements were fabricated by applying gallium focused-ion beam (Helios Nanolab 600i System, FEI Company) onto chemically synthesized single-crystalline gold flakes, as reported in our previous works [34-37]. The gold flakes were drop-casted on a cover glass coated with a 40 nm thick ITO layer. The ITO glass contains pre-fabricated markers for target identification. Linear Yagi-Uda nanoantennas, L-shaped Yagi-Uda nanoantennas and log-periodic nanoantenna were fabricated on four different gold flakes. Replicates of antenna elements were fabricated on the same flakes of the antenna for characterization of the optical response of individual elements. To facilitate optical measurement, the smallest distance between structures is set to be at least 1.5 μm. The width of all antenna elements is 50 nm, as determined by the fabrication pattern. The height is determined by the thickness of the flakes and the focused-ion beam milling process. After fabrication, the height of the fabricated YU-650, YU-850 and LPNAs is estimated according to SEM images to be around 40 nm and the height for LYUs is about 50 nm.

**2. Optical measurement**

Scattering spectra of nanostructures were acquired by a home-built dark-field microscope, where a unpolarized light from a broadband white light source (HAL 100 illuminator with quartz collector, Zeiss) illuminates the sample via an oil condenser (achromatic・aplanatic condenser, N.A. = 1.4, Zeiss). The scattered light was collected by an air objective (MPlanApo 60X air N.A. = 0.9, Olympus) on the opposite side of the condenser and aligned onto the entrance slit of a spectrometer (SR-303i-A with DU401A-BV CCD, Andor) with an achromatic lens (AC254-300-A-ML f = 300 mm, Thorlabs). A linear polarizer (LPVIS100-MP, 550-1500 nm, Thorlabs) was inserted before the achromatic lens to select the polarization of the scattering light. A circularly polarized 532 nm CW laser was used to excite one-photon photoluminescence from gold nanostructures. The excitation laser was focused from substrate side onto nanoantenna by an oil objective (PlanApo 60X Oil N.A. = 1.42, Olympus). The photoluminescence was collected by the same objective followed by a dichroic mirror and a notch filter for 532 nm to remove the laser scattering from the sample. Polarizer was inserted in the collection beam path to select the polarization. A confocal system was used to generate photoluminescence maps, which help to locate the feed element of nanoantennas for further selective excitation of specific antenna elements. For measurement of the PL spectra of single rods and gold octahedra, the excitation power was 0.10 mW and the integration time was 20 sec with gain set to 120. For spectral analysis of PL, the integration time for each spectrum was 30 seconds under averaged laser excitation power of 0.20 mW. Background spectra were measured on the substrate near the nanostructures and were subtracted from the corresponding photoluminescence spectra. Back-focal-plane images of photoluminescence were measured with an EM-CCD (iXon 897, Andor). For YU nanoantennas and L-shaped Yagi-Uda nanoantennas, the acquisition time has been set to 90 seconds with CCD gain value of 15 under laser excitation power of 0.30 mW. For log-periodic nanoantenna, the acquisition time for back-focal-plane images has been set to 120 seconds with CCD gain value of 30 under laser excitation power of 0.25 mW. The back-focal plane image of neighboring area of the nanoantenna was taken under same condition as background and subtracted from the back-focal plane image of the antennas. Band pass filter centered at 650 nm, 700 nm, 750 nm, 800 nm and 850 nm with a finite bandwidth of 40 nm (FKB-VIS-40, Thorlabs) were used to obtain photoluminescence back-focal plane image at different wavelength range. Figure S3 in the Supporting Information shows the schematics of the experimental setup.

**3. Simulation**

We have performed 3D full-wave numerical simulations using the finite-difference time-domain method (FDTD Solutions, Lumerical Solutions) to simulate the scattering spectrum of isolated nanoantenna elements and far-field radiation patterns. The structural dimensions used in the simulations were estimated according to the SEM images of the fabricated nanostructures. The 40-nm thick ITO layer has been considered in the simulations. To simulate far-field scattering spectrum, a total-field scattered-field plane wave source has been used to excite the antenna from the glass half space. All perfectly matched layer (PML) boundaries of the simulation volume were set to be 3,000 nm away from the antenna center to avoid incorrect absorption of the near fields by the PML. Mesh step for space discretization in x, y, and z directions has been set to 2 nm within the volume covering the antenna structure. The scattered power was obtained by integrating the Poynting vector over the surface of a closed box consisting of six two-dimensional power monitors enclosing the nanoantenna. The transmitted power was normalized to the total source power. To simulate radiation pattern of the photoluminescence emission from the feed element, broadband dipole source was placed in the middle of the feed elements with polarization parallel to the longitudinal direction. Mesh step of the space discretization in x, y, and z directions was set to be 3 nm within the volume covering the antenna. The far-field radiation from nanoantenna was recorded by a 2D monitor (20000 nm × 20000 nm) placed in the substrate. This large 2D monitor was necessary to collect radiation over large emission angle. To simulate the experimental emission pattern taken with a polarizer, back-focal plane images are obtained by selecting the field component along the longitudinal direction of the antenna elements. The simulation settings are illustrated in Fig. S4 in the Supporting Information.

# - Supporting Information -

## Photoluminescence-driven Broadband Transmitting Directional Optical Nanoantennas

*Kel-Meng See[1], Fan-Cheng Lin[1], Tzu-Yu Chen[1], You-Xin Huang[1], Chen-Hsien Huang[1], A. T. Mina Yesilyurt[2] and Jer-Shing Huang[1,2,3,4*]*

[1] Department of Chemistry, National Tsing Hua University, Hsinchu 30013, Taiwan

[2] Leibniz Institute of Photonic Technology, Albert-Einstein Str. 9, 07745 Jena, Germany

[3] Research Center for Applied Sciences, Academia Sinica, 128 Sec. 2, Academia Road, Nankang District, Taipei 11529, Taiwan

[4] Department of Electrophysics, National Chiao Tung University, Hsinchu 30010, Taiwan

Correspondence should be addressed:

jer-shing.huang@leibniz-ipht.de

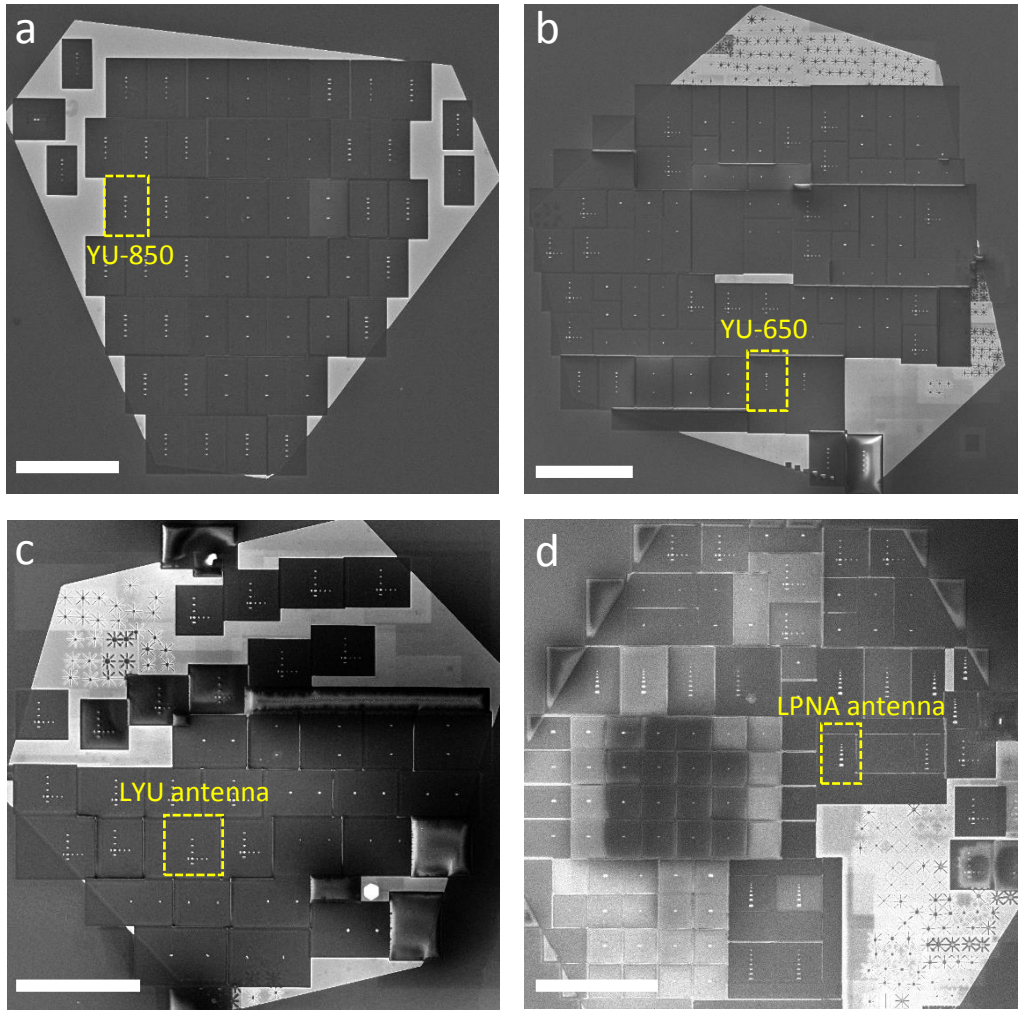

**Figure S1.** a. SEM image of the gold flake milled by focused-ion beam to fabricate linear Yagi-Uda (YU) nanoantennas designed for directional emission at 850 nm. Individual feed, reflector and director elements were also fabricated on the same flake. The indicated structure is the one shown in the Fig. 2a in the main text. b. SEM image of the gold flake milled by focused-ion beam to fabricate linear YU nanoantennas designed for directional emission at 650 nm. Individual feed, reflector and director elements as well as tester structure of L-shaped Yagi-Uda nanoantennas (LYU) were also fabricated on the same flake. The indicated structure is the one shown in the Fig. 2a in the main text. c. SEM image of the gold flake milled by focused-ion beam to fabricate LYU nanoantennas designed for directional emission at 850 nm to the north (top) and 700 nm to the east (right). The indicated structure is the one used in the Fig. 3 in the main text. Individual feed, reflector and director elements were also fabricated on the same flake. d. SEM image of the gold flake milled by focused-ion beam to fabricate Log-periodic nanoantennas (LPNAs) designed for broadband directional emission between 650 nm to 850 nm. Individual feed, reflector and director elements were also fabricated on the same flake. The indicated structure is the one shown in the Fig. 4 in the main text. All scale bars are 5 micrometers.

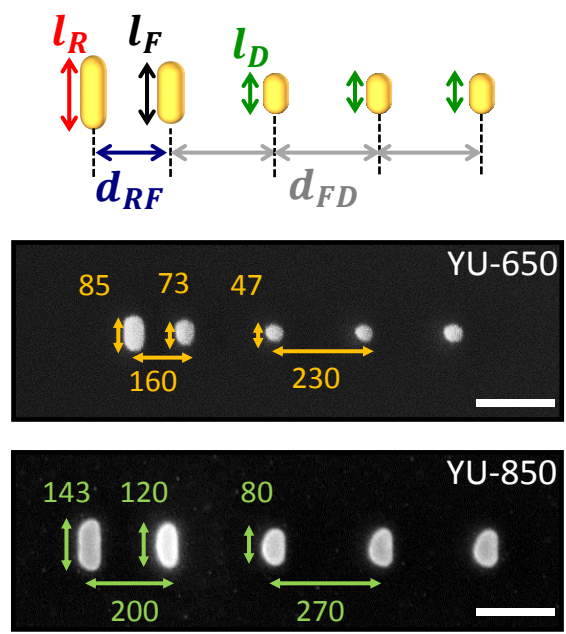

**Figure S2.** Design and dimensions of the fabricated Yagi-Uda nanoantennas (top). The width and height of the elements are 50 nm and 40 nm, respectively. SEM images of the linear YU nanoantennas designed for directional emission at 650 nm (YU-650, middle) and 850 nm (YU-850, bottom). Dimensions are indicated with double arrows and numbers in unit of nanometer. Scale bars are 200 nm.

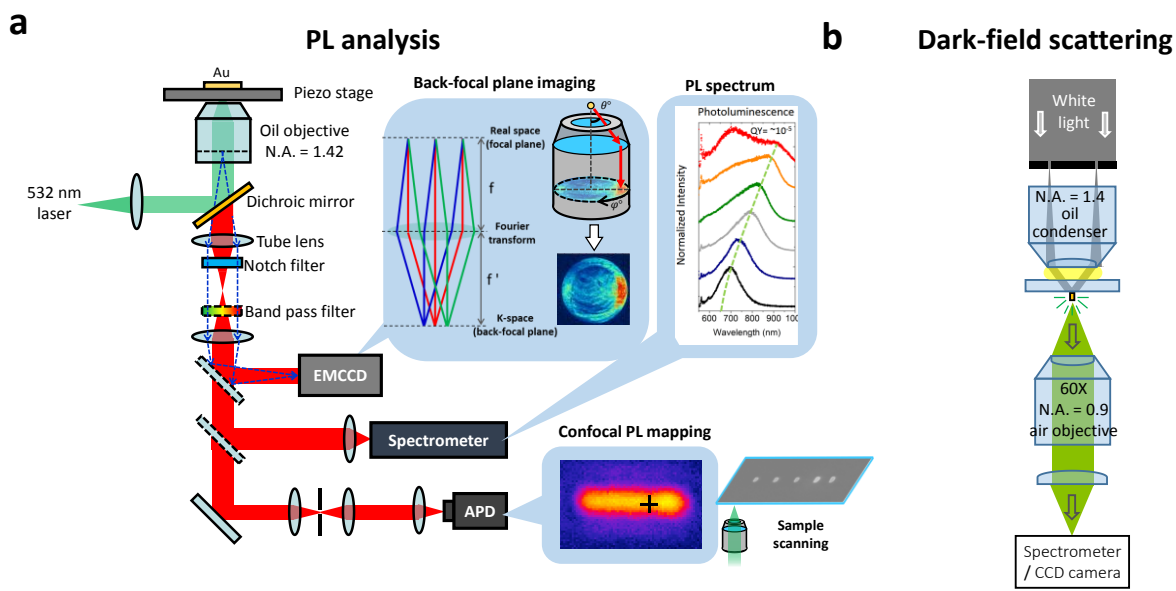

**Figure S3. a.** Optical setup for PL analysis, which includes three major functions, namely PL mapping, PL spectral analysis and back-focal plane imaging. **b.** Optical setup for acquiring dark-field scattering spectrum of nanostructures.

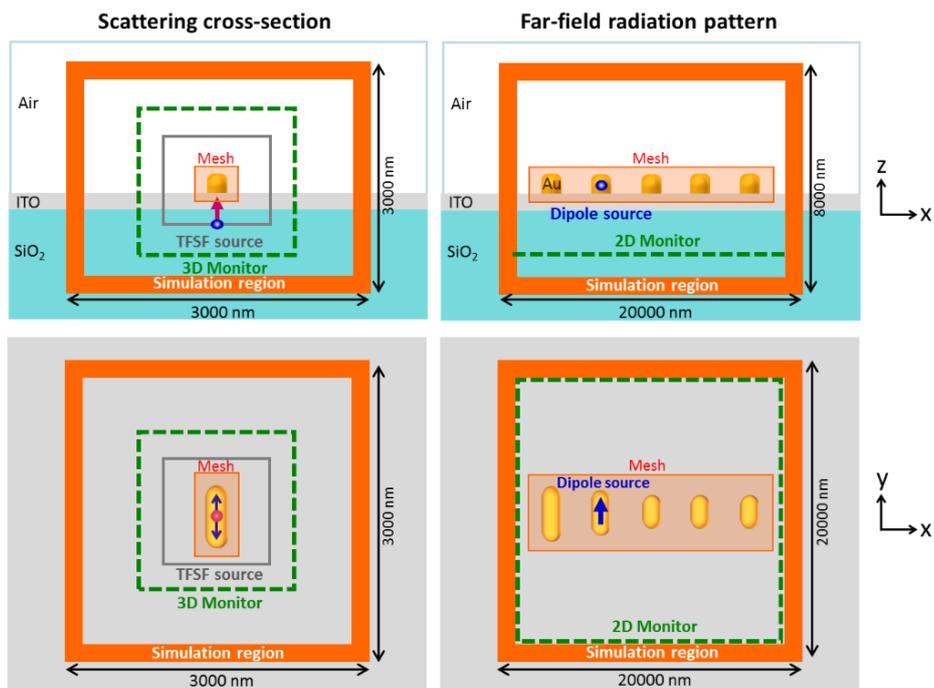

**Figure S4.** Schematic illustrations of the settings in the FDTD simulation. Note that for better illustration, the dimensions and distances do not reflect the real scale.